# The Unfolding and Control of Network Cascades

Adilson E. Motter[1,2] and Yang Yang[1,3]

[1]Department of Physics and Astronomy, Northwestern University, Evanston, IL 60208, USA.
[2]Northwestern Institute on Complex Systems, Northwestern University, Evanston, IL 60208, USA.
[3]Department of Chemical and Biological Engineering, Northwestern University, Evanston, IL 60208, USA.

*In network systems, the same connections that provide functionality can also serve as conduits for the spread of failures and innovations that would otherwise be limited to small areas.*

It is just a small exaggeration to say that we have reached the "network age." Now both more systems are recognized as networks—think of engineered materials, intracellular media, organismal physiology, ecological systems, swarming robots—and more networks have come into existence due to human activity—global financial and transportation networks; continental-wide power grids; the Internet; and large-scale social, communication, and information networks (Adilson E. Motter and Réka Albert, Physics Today, April 2012, page 43). A characteristic property of networks is their ability to propagate influences, such as infectious diseases, behavioral changes, and failures. An especially important class of such contagious dynamics is that of cascading processes. These processes include, for example, cascading failures in infrastructure systems, where a network component may fail or disable itself in response to the failure of other components; extinctions cascades in food-web networks, where perturbation to the population of one species may cause a sequence of others to undergo extinction; and information cascades in social systems, where an individual's behavior is influenced by the behavior of others (Fig. 1, Box).

Cascading processes have intrinsic features that distinguish them from the simple contact processes that typically characterize other network-spreading phenomena like epidemic spreading and diffusion—which we can illustrate by comparing failures in a power grid with a model for the spread of flu in an unimmunized population of like individuals. First, the likelihood of a node failing increases non-additively as a function of the number of other failures in the neighborhood, whereas in epidemic spreading the probability of acquiring the disease from a neighboring node does not depend on the state of other nodes. For example, a power station may never fail if only one connected station has failed, whereas an individual has a nonzero probability of contracting flu even if only one contact has the virus. Second, the propagation of cascading failures is not restricted to being local, in the sense that a power station may fail even if none of its close neighbors have failed, whereas in the epidemic case the virus can only reach an individual through a neighbor that is contaminated (assuming the network is the only medium for the transmission). Third, the impact of one power station failure on other individual stations can be disproportionally large compared to the average, whereas, all other factors being the same, the transmission probability is expected to vary little across different flu-infected individuals. Each of these properties, although illustrated here for cascading failures in power grids, also applies to most other cascading processes.

These differences have major implications. The *non-additivity* means that cascades can be more likely to propagate in networks with structures that allow reinforcement from multiple neighboring nodes (or connections), such as locally redundant networks despite their larger average node-to-node distance [1], whereas epidemics propagate more efficiently in networks with long-range connections, such as random ones. The success of viral marketing and information sharing through social media, for example, may depend critically on the interplay between network structure and the extent to which this property underlies the dynamics. The *non-locality* indicates that the state change of a node may remotely change the state of other nodes without changing the state of intermediate nodes [2], which has no analog in epidemics. This property is expected to be more pronounced in networks with small clustering and large average path length, like power grids. The *disproportional impact* means that the influence of the change of state of a node depends not only on the connectivity pattern of the node—instead, the node itself may be more influential. The latter relates to the important issue of intrinsic fitness versus position in the network, which has implications for cascades as well as numerous other processes.

Cascades are generally associated with new initiating events rather than equilibrium states or an entity that has been persistent in the network, whereas a contagious disease can remain endemic in a nonhomogeneous population even at low effective spreading rates. For example, in 1970 gonorrhea led the list of infectious diseases in the United States despite evidence that most infected individuals would transmit the disease to less then one partner on average. The proposed explanation recognized the existence of a core (potentially asymptomatic) portion of the population with higher infection rate; it was estimated that a core of only 2% of the susceptible individuals would be responsible for 60% of all infections [3]. Now, had this been a cascade, the need of reinforcement would effectively limit the infected population to the core and nothing else.

These distinctive properties have ramifications for the modeling, detection, and control of cascade dynamics. Incidentally, while here we refer to nodes or connections, failure or adoption, etc. for concreteness, in the most general case a cascade can involve a variety of status changes in any or multiple types of network components. Accordingly, for many examples given, similar conclusions also apply to different forms of cascades.

*Box – Examples of cascades*

Cascades are self-amplifying processes that may cause a substantial part of the system to change behavior in response to a relatively small event. These attributes render cascades a threat in systems in which failures can propagate through the network, such as in blackout-causing cascading failures in power grids, congestions in traffic networks, cascades of secondary delays in air transportation, default cascades in financial networks, disruption cascades in supply networks, diseases caused by malfunction of biochemical pathways, and extinction cascades in ecosystems. Cascades also underlie the operation of numerous systems, with examples including biochemical cascades in signaling networks, plea bargain cascades in criminal justice systems, and cascades of technology-adoption and cooperation in contributing to the public good in society. In certain scenarios, cascades of interest can be designed, such as in get-out-the-vote and other behavior adoption campaigns, in re-sharing of advertising content on social media, including Facebook and Twitter, and in viral marketing that influences purchasing decisions on products.

**Cascade models**

An initial step in the study of network cascades is the mathematical representation of the process. This step, as it turns out, is nontrivial because while spread—and hence self-amplification—is an easy feature to capture, non-additivity, non-locality, and disproportional impact are not. Accordingly, the compromise between being simple enough to be amenable to analysis and being comprehensive enough to represent reality is hard to come by for models of cascade dynamics. Significant progress has been made, nevertheless, by tailoring models to the research question at hand. Some of the most widely studied cascade models can be grouped into the following (not mutually exclusive) classes.

*Avalanche models*. These are models inspired by the Bak-Tang-Wiesenfeld sandpile model in which a grain is placed on a random node at each time step and nodes topple, distributing their grains (part or all) to the neighbors, when a threshold is achieved (Fig. 2a). Thus, following one such event, the network may experience no further toppling or a toppling cascade that impacts multiple nodes. The thresholds can be heterogeneous across the network, which may lead to disproportional impact; avalanche models immediately account for non-additivity, but, like other first-neighbor models, they do not allow for non-locality [4, 5].

*Percolation-related models*. In these models, the end question usually is whether the set of nodes that change state percolate, in the sense of forming a large-scale connected cluster in the network (comprising a nonzero, or "macroscopic," portion of all nodes in the infinite network size limit). For this reason, even though the specifics of the dynamics may vary from model to model, the solution of the model is usually reduced to a percolation problem, which is purely structural. An example is the threshold model [6], used to study many types of cascades in social systems, in which the state of a node will change when a threshold fraction of the neighbors have changed (Fig. 2b). Notice that this model does not account for disproportional impact or non-locality, although it accounts for non-additivity. The solution of this model in random networks shows, for example, that increasing the heterogeneity of the distribution of connections per node makes the system less susceptible to large-scale cascades but increasing the heterogeneity of the thresholds makes it more susceptible [7]. While reducible to percolation problems, these models should not be confused with percolation processes used to study the structural robustness of networks, which do not account for the dependent failures that often correspond to the lion share of all failures in a cascade.

*Statistical models*. These are models that do not attempt to account for the detailed dynamics or even structure of the network, but are instead designed to describe the expected number of network elements joining the cascade in each generation of a time-discretized cascade event. A prime example of such bulk models is branching process models (Fig. 2c), which in their simplest form are characterized by a main parameter λ representing the average number of offspring failures for each parent failure in the previous generation (see, e.g., [8]). Because these models do not specify the causal relations between parent and offspring failures, nor the identity of the elements involved in the cascade, they do not take a stand on any of the properties unique to cascades. They can permit, however, the statistical prediction of the outcomes, including cascade size, based on simulated or historical data.

*Flow redistribution models*. Many real networks exhibiting cascades (traffic, electric, metabolic networks) are in fact flow networks. In models designed to capture this feature, the dynamics is such that when a network component is inactivated, the flow through that component is redistributed over other nodes and connections, potentially reaching the capacity of those elements (Fig. 2d). The capacity can be a hard or a soft constraint and may imply complete, partial, or delayed failure, or just saturation. In any case, a failure can lead to further flow redistributions and failures. A simple model incorporating these elements is one in which the flow exchanges between any two nodes are transmitted through the shortest paths connecting them [9]. Despite its simplicity, this model captures all three salient properties of cascades singled out above. This model also reproduces the robust-yet-fragile property of networks, in that they tend to be robust to perturbations on any of the many low flow components but fragile when one of the few high flow components is perturbed. System-specific realistic models have also been considered, of which we highlight the power-grid cascading failure models based on the simulations of power flow equations [10].

In contrast with the models above, which may involve *ad hoc* assumptions tailored to reproduce cascades, there are two broad classes of models that can be derived directly from representations of the dynamics in the network and that reproduce the cascades as special manifestations of that dynamics.

*Dynamical-systems models*. These are models directly derived from the dynamical equations describing the state of the network, typically in the form of a large set of coupled ordinary differential equations. For example, in a food-web network these could be the consumer resource equations, and in a network of power-grid generators they could be the swing equations that follow from Newton's second law. These systems are nonlinear, dissipative, and generally multi-stable, with some of their stable states (or attractors) representing desired states (e.g., states in which all species coexist) and others representing undesired states (e.g., states with one or multiple extinctions). In this "continuous" description, a cascade is then interpreted as the process through which the system is driven by a perturbation from the attraction basin of a desired state into the attraction basin of an undesired one, and then evolves towards that undesired state [11] (Fig. 3a). One of the benefits of such models is that, in addition to immediately accounting for all defining properties of the cascade dynamics, they also allow the study of numerous implications. For example, in the presence of sequential perturbations, it follows immediately that the outcome will depend strongly on the order and timing of the perturbations. This has been illustrated for extinction cascades, where, depending on perturbation scheduling, a cascade triggered by the suppression of one species often can be enhanced, inhibited, or completely mitigated by the deliberate suppression of other species [12].

*Agent-based models.* These are models that are implemented using agent-based modeling, which is a computational framework in which the system is represented as a collection of interacting agents and simulated at the individual rather than aggregate level. Each agent, which may represent a node in a network, responds to the behavior of neighboring nodes in this network according to pre-defined rules [13] (Fig. 3b). Agent-based models can allow fairly realistic representation because they are applicable even when so many details need to be included that it is no longer practical to represent the system in terms of closed-form equations. They also naturally account for possible time dependencies in the network of interactions between agents. These models can be used to study cascading processes in a variety of contexts, ranging from power engineering and molecular biology to economics.

Another way in which cascading models can be classified is as abstract models, simplified models, and detailed models. In a power grid [10], for example, the detailed models would designate the most realistic, causal representation of the system, the simplified models would involve some conscientious approximations (e.g., the use of DC power flow approximation), and abstract models would not directly account for the physics of the power flows but instead focus on implications, such as in applications of branching models. Each model has its own advantages and limitations, and the suitability of a model will largely depend on the research question under consideration.

**Early detection and prediction of cascades**

Proper response to a cascade event requires the prediction of cascades at the time they are triggered or early detection at the time they start to propagate. While it is in principle straightforward to determine when a network component changes state, the system-wide impact of such a change is not, which by itself sets constraints on one's ability to determine the potential of a given condition to develop further into a cascade. There are therefore essentially two main scenarios under which cascades can be predicted: 1) the dynamics is known and can be simulated accurately given the observed state of the network; 2) data is available from previous events under similar conditions and can be used to statistically infer the outcome. A challenge with 1) is that each cascade involves a potentially different complex sequence of dependent changes of state, which can be sensitive to modeling and state estimation uncertainties (much in the same way as the long-term predictability in systems as simple as a double-pendulum can be compromised by small uncertainties [Adilson E. Motter and David K. Campbell, Physics Today, May 2013, page 27]). A difficulty with 2) is that large cascades, which are precisely the most relevant ones, are rare events for which the statistics of previous events tends to be inherently limited.

Despite these challenges, several approaches have been developed. In such approaches, one may be interested in a binary answer (e.g., whether or not a cascade is likely), a continuous answer (the probability of a cascade as a function of cascade size), or the full description of the cascade trajectory. In the latter case, one may seek to specify not only the size but also the time evolution of the cascade and identities of the network components involved.

A first step in many data based approaches is to identify *features* that are strongly correlated with specific cascade outcomes. For example, in cascades of information sharing on Weibo/Twitter and Facebook, the activity of key users as well as the temporal and structural properties of early-sharing events are all strongly correlated with the occurrence of large cascades [14]. Making use of these features, some approaches employ machine-learning techniques that can systematically account for multiple such correlations, such as logistic regression, neural networks, and support vector machine, to predict the most likely outcomes. Naturally, in the *real-time* detection of cascades, a potentially limiting factor for many approaches is the time required for collecting and processing the data associated with sensing the state of the network elements, which has to be smaller than the time scale for the propagation of the cascade itself.

It is also useful to consider the prediction of cascades using models. On the one hand, properly validated models that can predict the dynamics of the system fairly realistically, such as the TRELSS model to evaluate power network reliability [10], can be simulated in tandem with the observation of the real system to predict a cascade given a detected perturbation. The use of such models is fundamentally different from statistical and classification approaches as they seek to forecast the dynamics of the system and hence do not rely on data from previous cascades. On the other hand, it is conceptually instructive to identify within models the key factors involved in giving rise to cascades. At the most elementary level one can simply apply approaches such as the ones in the previous paragraph, to simulated data rather than empirical data—for example, for the overload model in [9], it follows that cascades are more likely to be triggered by the failure of nodes with high centrality. Sensitivity analysis can be used to investigate variants of the latter in other flow networks (Fig. 4).

In the study of models, a more complete characterization is also possible in many cases. In the threshold model on random networks [7], for example, the activation of a single node will trigger a large-scale cascade if and only if the node is connected to a large-scale (i.e., "percolating") cluster of early adopters (nodes that became active whenever one neighbor becomes active); more profoundly, in sparsely connected networks one such cascade will include a large-scale portion of *non*-early adopters. In dynamical-systems models where the non-cascading state is represented as a stable state, a perturbation will generate a cascade if and only if it brings the system outside the attraction basin of this state [11]; moreover, since the entire trajectory of the cascade can be determined by evolving the trajectory of the system, the final state of the cascade can be predicted in simulations. Incidentally, this continuous description shows that it is a simplification to think of a cascading failure as the process in which one failure leads to another. Instead, it is the continuous change in the full state of all variables that drags the system into the path of successive failures—an observation that has practical implications for early cascade detection, as it shows that it is not enough to track the component failures. In a power grid, for example, proper intervention requires the cascade to be detected before secondary failures occur—a nontrivial problem closely related to the problem of accurately assessing the state of the network.

**Mitigation, enhancement, and control of cascades**

Whether to put in motion a cascade of interest or to stop a detrimental one, cascade control is a primary ambition in the study of cascade dynamics. A convenient way to think about cascade control is by considering the manipulation of *cascade risk*, defined as the product between the probability of cascade occurrence and cascade cost. Risk adequately characterizes, for example, the scenario in which the suppression of frequent cascades inadvertently increases the likelihood of rare but very large ones. In the case of beneficial cascades, the counterpart of risk is the probability times the payoff—also known as *expected utility*. More generally, such objective functions can be interpreted as "probability" x "size" for a properly defined notion of cascade size.

The opportunities for control may be different depending on whether the goal is to inhibit or promote cascades, whether the system is engineered or natural, and the extent to which individual network elements can be "actuated" before and after a cascade is initiated. In social systems, a conceptual starting point for the launching of a successful (large) cascade of information sharing, technology adoption, or behavioral change, is the notion that some individuals (or organizations) are significantly more influential than others—known as the *influentials hypothesis* [15]. The concept is similar in spirit to disproportional impact, where influence may depend as much on network position as on reputation and other aspects of intrinsic fitness. A competing hypothesis, supported by the threshold model, is that there are no influentials and that the ability to trigger cascades depends much more on the structure of the network (e.g., the subnetwork of early adopters) [15]. Studies focused on maximizing the spread of influence suggest the existence of influentials in some models [16]. While the extent to which influentials are influential may remain controversial, the idea is that even if they have a higher threshold for joining the cascade, once they do they are more likely to cause others to join, making them ideal targets for early adoption. Needless to say, another control parameter in this case is the appeal of the subject of the cascade (the information, technology, or behavior).

In engineered networks, on the other hand, the most immediate action toward preventing cascading failures would be to design the system to be resilient, with components that are not prone to fail and/or that can withstand the most likely disturbances, or to be robust against the failure of part of their components. In a related body of work, much has been learned from research focused on identifying network properties such as redundancy and community structures (see, e.g., [5]) that have the potential to reduce the likelihood of cascading failures. A complementary layer of protection comes from proper management of the system. A power grid, for example, is equipped with controllers at multiple levels to keep the system balanced under normal operating conditions [Scott Backhaus and Michael Chertkov, Physics Today, May 2013, page 42], which combined with proactive operational decisions (e.g., adjustment of capacity reserve or load level as conditions change) are key to reduce risk. However, while a necessary step, in practice no engineering effort can completely eliminate the possibility of occasional failures and hence of the occasional triggering of cascades in large complex networks (and system redesign would not always be an option in natural and resource-limited networks).

There is therefore interest in what could be perceived as the most ambitious form of cascade control: controlling unforeseen or unavoidable cascades *after* they have been triggered. Still using the power grid as a model system, consider a disturbance that either causes the capacity of some components to be violated and/or an imbalance between power demand and supply. With the goal of rebalancing the system to prevent the propagation of the disturbance while minimizing the amount of power lost (difference between power demand and power supplied), some of the main control actions that can be taken are the rerouting of power flows, load shedding, and dispatch of power generation. This is a simplification, of course, as there are additional factors to consider, such as time scales, other instabilities, cost of power generation, and priority-circuits (e.g., hospitals vs. street light). In any case, there are basically two main types of approaches to address the problem of minimizing power loss: centralized algorithms, which require state information from the entire system and can in principle deliver a globally optimal solution, and decentralized ones, which rely on local state information and can generate locally optimal solutions that are computationally less expensive. The approach that is most adequate will depend largely on the scale of the system under consideration. The ongoing development of smart grids will lead to a number of new possibilities, such as real-time pricing and large-scale use of smart appliances, which will allow control not only of supply and delivery but also of demand. The latter may prove crucial to increase the penetration of renewables from intermittent sources.

Dynamical-systems models offer a particularly insightful context for the post-triggering control of cascades. In these models, the triggering of a cascade is associated with an event that brings the system outside the attraction basin of the desired attractor. The goal is to steer the trajectory back. In deterministic systems there are essentially two ways in which that can be done: by perturbing the system variables to bring the state to the desired basin (Fig. 5a) or by perturbing the system parameters to bring the desired basin to the state (Fig. 5b)—the extreme of which is a bifurcation that leaks the occupied basin into the other. Using the game of billiard to build an analogy, these two approaches would be akin to pocketing the balls by striking then with a cue versus by tilting the table itself. The challenge in the control problem is that constraints on feasible interventions limit the directions and extent to which one can move in state and parameter space; and determination of the global structure of the basin boundaries (or quasipotential) of the system, which would be required in simple control approaches, is computationally impossible in the typically high-dimensional dynamics of large networks. Two approaches have been developed recently that overcome these challenges. One approach locates the basin of the target attractor without any *a priori* information about its location [11]. The other approach manipulates the height of the boundary along the least action path connecting the attractors to induce a bifurcation that eliminates the undesired attractor while not changing the stability of other states in the system [17]. Within the dynamical-systems modeling, in principle one can go beyond suppressing or enhancing cascades to instead control the entire cascade trajectory—which can be relevant to avoid adverse effects.

**Current and future research**

Cascading failures are prominent in the realm of possible threats to a network because global effects do not require global actions: the network is disrupted not by the disturbance itself but by the chain of events it puts in motion. Like in the case of other spreading phenomena, traditionally more effort had been put into the analysis than into the prediction and detection of cascades. Yet, recent advances in the latter combined with ongoing advances in cascade control are now creating the possibility of real-time control of cascades already in course. Another timely topic concerns the post-cascade dynamics and, in particular, the nontrivial matter of restoring the system after a cascading failure.

More broadly, essentially every problem discussed here is still work–in-progress around which significant research activities continue to take place. This includes the discovery of new mechanisms underlying cascades, relations between network structure and cascade dynamics, trade-off between resilience to frequent versus large cascades, control in the presence of uncertainty, characterization of the extent to which the trajectories of different cascades are different, identification of common features across different domains, experimental validation of longstanding hypotheses, and development of new applications that leverage cascades—just to name a few.

The above applies not only to systems modeled as a single network with a single type of node and a single type of interaction, but also to many networks composed of multiple types of nodes and interactions. In social networks, for instance, individuals may be distinguished by gender, age, profession, and other characteristics; their interactions may be physical or virtual, personal or professional, etc. The same applies to many other complex systems. A city, for example, can be regarded as a network that includes infrastructural, social, economic, and biophysical network layers that interact with one another. While most of the discussion carried out this far applies to such systems when regarded as one network, even if a heterogeneous one, it is sometimes convenient to separate the network layers of different nature and regard the system as a network of interacting networks (also referred to by a number of other terms, such as multilayer and multirelational networks). Additional questions about cascades arise when adopting this description, such as the differences (if any) from the case in which each component network is assumed to operate in isolation.

Taking for concreteness the case of two networks of *N* nodes each, coupling between the two networks creates the possibility that a cascade initiated in one network will propagate to the other network, hence leading to a larger cascade than in a single network. This does not mean, of course, that the cascade will be any larger than it would be in a single network of *2N* nodes. The coupling between two networks may also reduce the risk of cascades, as shown for a model network consisting of similar sub-networks coupled together [5], and as expected when one network stabilizes the other (as is the case for regulatory or control networks). Interdependent networks (a special type of interacting networks) serve as another instructive example. In the simplest form of such networks, each node has to be connected to nodes from different networks in order to be functional. It has been shown that, depending on the details of the model and network structure, interdependence may make cascades and percolation transitions both more and less abrupt [18]. The impact of interactions between networks is therefore far from obvious, likely system-specific, and yet to be fully understood.

The consequences of a cascade in a network can be undesired or desired, depending on the system, on the subject of the cascade, and on the beholder. The understanding, modeling, and control of cascades have been longstanding problems of practical importance in various application areas, which have nevertheless been successfully advanced by recent research in network dynamics. We suggest that the solution to pressing problems as diverse as the development of self-healing networks, the discovery of new therapeutic treatments, the prevention of widespread financial crises, and the design of effective behavior change campaigns, will all directly benefit from future advances in this rapidly developing multidisciplinary field.

**Acknowledgement:** The authors acknowledge support from the National Science Foundation (Grant No. DMS-1057128).

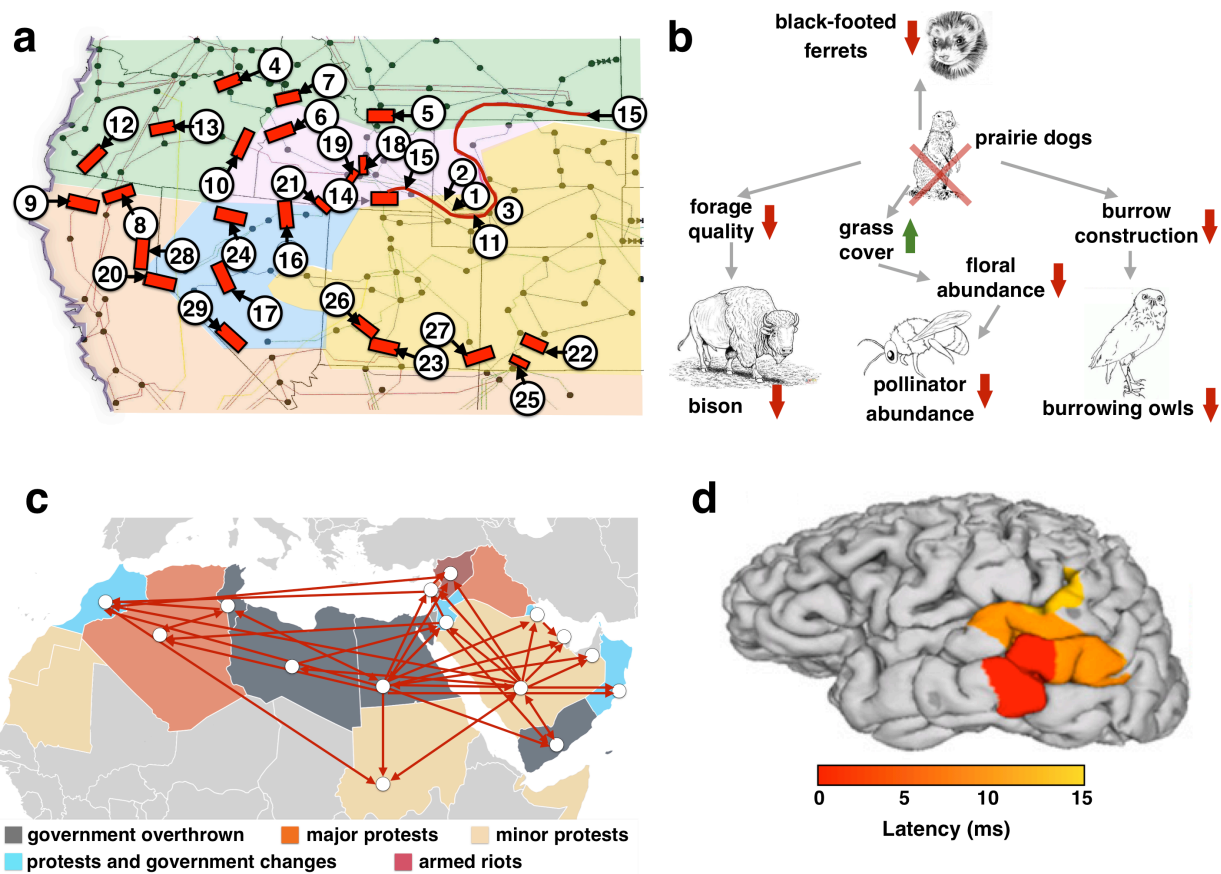

**Figure 1. Examples of cascades in networks.** (a) July 2, 1996 cascading blackout in the Western North American grid, in which a sequence of line outages (red bars) separated the network into five islands (colored areas), affecting more than two million customers (adapted from the WECC Disturbance Report). (b) Cascade of population declines caused by the loss of the black-tailed prairie dog in the USA's central grassland, which led to the decline (red arrows) in predators and other species that depend on the habitat that they create (adapted from B.J. Bergstrom *et al.*, Conserv. Lett. **7**, 131 (2014)). (c) Social cascade underlying the 2010-12 Arab Spring protests, where the dominant Facebook friendship connections are shown as a proxy of the influence across counties (adapted from C.D. Brummitt *et al.*, J. R. Soc. Interface **12**, 20150712 (2015)). (d) Example of neuronal avalanche color-coded by latency, where a cascade of bursts of neuronal activity propagates in the brain (adapted from J.M. Palva *et al.*, Proc. Natl. Acad. Sci. USA **110**, 3585 (2013)).

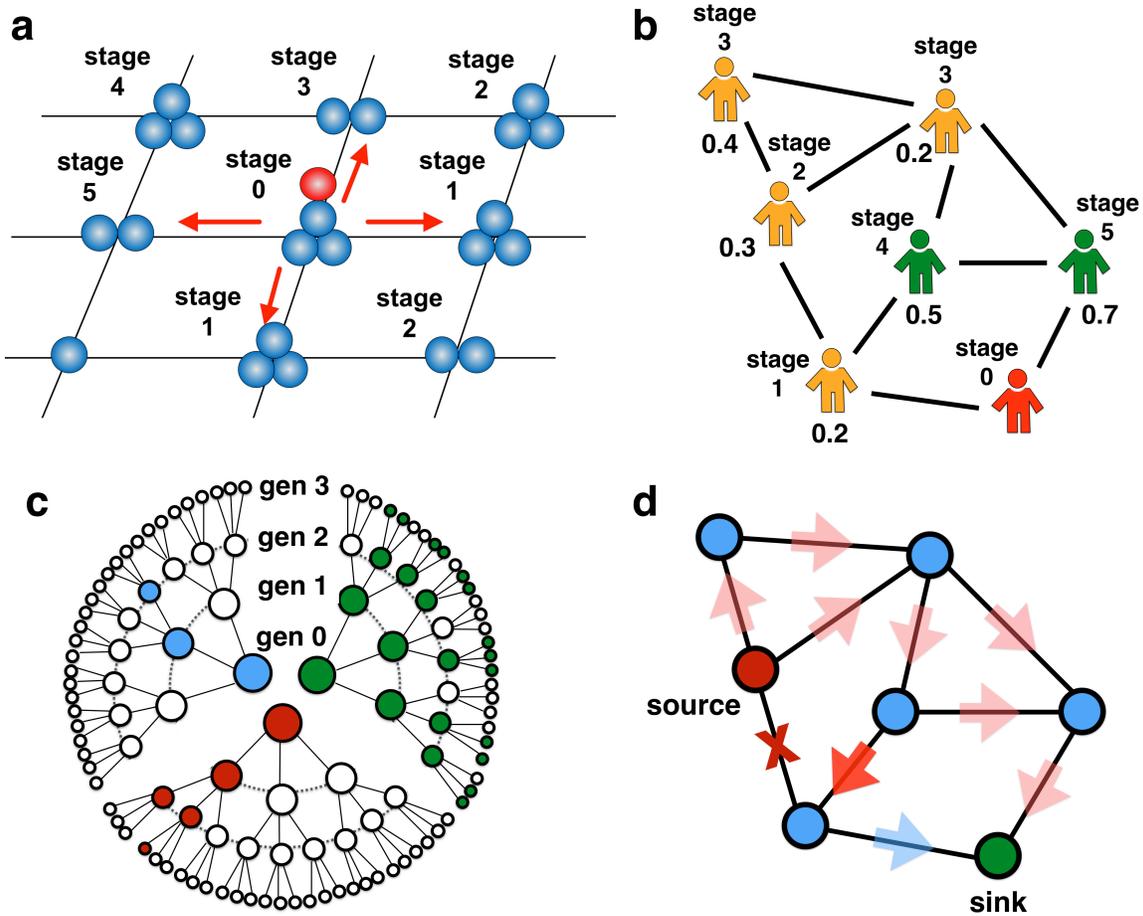

**Figure 2. Schematic representation of cascade models.** (a) Sandpile model on a lattice, where the center node reaches the threshold for toppling (assumed to be four). (b) Threshold model for an example assignment of adoption threshold (marked next to the nodes). (c) Branching-process model (in which the maximum number of offspring failures per parent failure is three), for cases where the number of new failures in each generation (gen) is larger (green), smaller (blue), and equal (red) to the critical number for cascade growth (adapted from T.P. Vogels, *et al*. Annu. Rev. Neurosci. **28**, 357 (2005)). (d) Flow-redistribution model, where the inactivation of one connection (cross sign) causes the increase of flow through other connections (red).

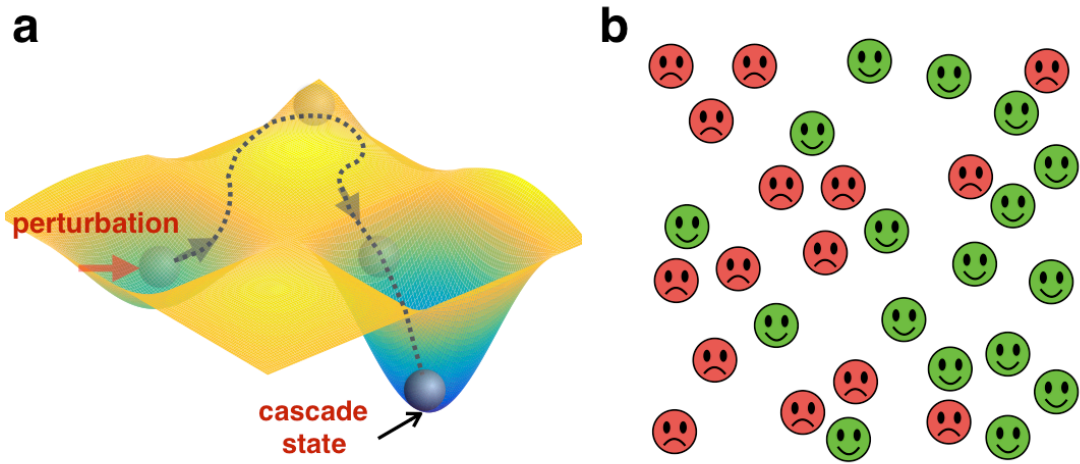

**Figure 3. Schematic representation of cascade models (continued).** (a) Dynamical-systems model, where a perturbation (red arrow) can drive the system towards an attractor representing a cascade state. (b) Agent-based model, where at each time step each agent follows pre-defined rules to update its state (red or green) on the basis of the states of others.

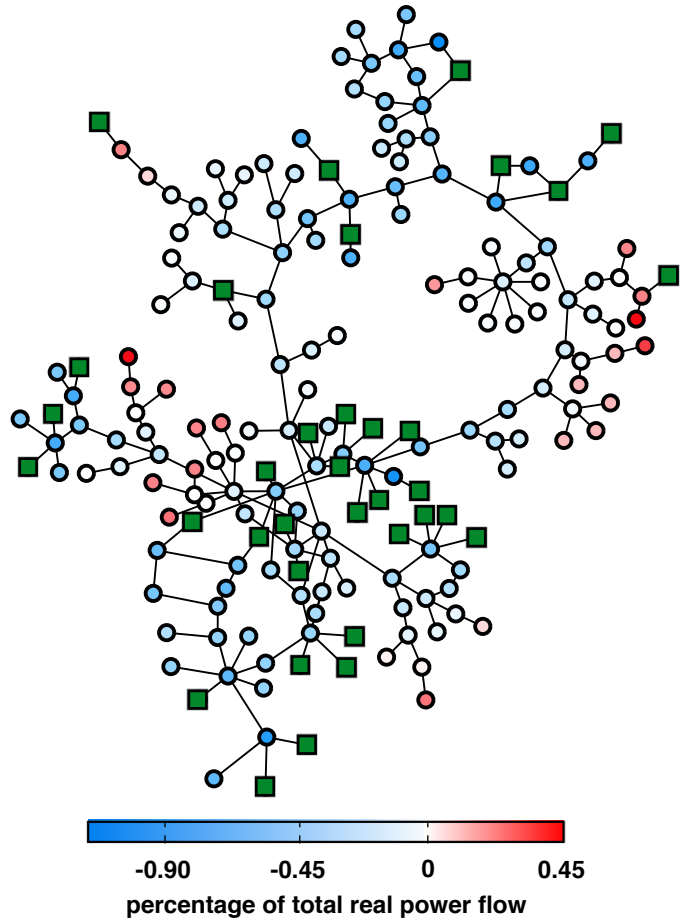

**Figure 4. Sensitivity map of the Iceland power network**, where the color on each load node indicates the change in the aggregated flow over all power lines as power demand on that node is increased by 1%. The percentage is relative to the total power generated in the network. To keep the balance of supply and demand, power demand is decreased uniformly across all other load nodes. Loads are indicated by circles, while squares represent generators.

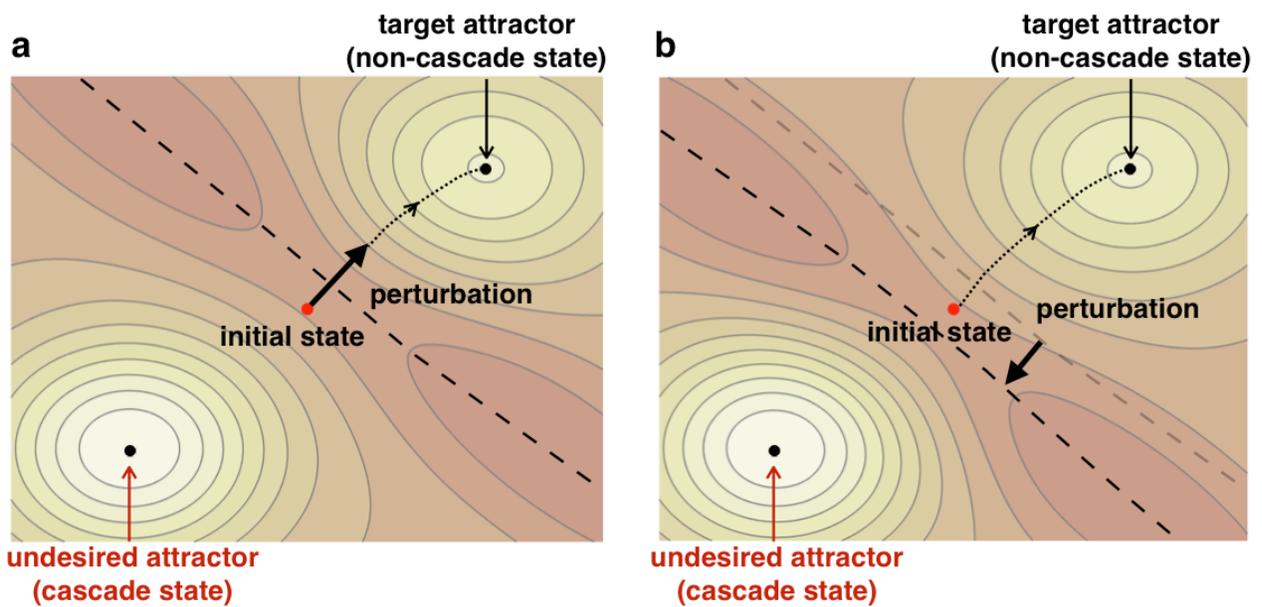

**Figure 5. Control of cascades in a dynamical-systems description**, where the initial state is in the basin of an undesired attractor. A system on the verge of cascading can be nudged back toward a target attractor by perturbing either (a) the state (red dot) to move the system toward the target basin or (b) tunable parameters to shift the boundary between the desired and undesired basins of attraction (dashed line). Lighter colors represent deeper points in the attraction basins, and contours are lines of equal depth. (Adapted from refs. [11] and [17].)